\documentclass[referee]{chjaa}          

\usepackage{graphicx,times}             
\usepackage{lscape}
\input{epsf.sty}                        
\input{psfig.sty}                       

\begin{document}

   \title{Spectral study on the dips of Cir X-1
}

   \volnopage{Vol.0 (200x) No.0, 000--000}      
   \setcounter{page}{1}           

   \author{Ya-Juan Lei
      \inst{1}\mailto{}
   \and Fang-Jun Lu
      \inst{1}
   \and Jin-Lu Qu
      \inst{1}
   \and Li-Ming Song
      \inst{1}
   \and Cheng-Min Zhang
      \inst{2}
      }

   \institute{Laboratory for Particle Astrophysics, Institute of
 High Energy Physics, Chinese Academy of Sciences, Beijing, 100049, China\\
             \email{leiyj@mail.ihep.ac.cn}
        \and
         National Astronomical Observatories, Chinese Academy of Sciences, Beijing 100012, China   \\
        }

   \date{Received~~2007 month day; accepted~~2007~~month day}

\abstract{
We present X-ray spectral analyses of low mass X-ray binary Cir X-1 during X-ray dips, 
using the {\sl Rossi X-ray Timing Explorer (RXTE)} data.
Each dip was divided into several
segments, and the spectrum of each segment was fitted with a three-component blackbody
model, in which two components are affected by partial covering and the third one 
is unaffected. A Gaussian emission line is also included in the spectral model to 
represent the Fe K$\alpha$ line at $\sim$ 6.4\,keV. The fitted temperatures of  
the two partially covered 
components are about 2\,keV and 1\,keV, while the uncovered component
has a temperature of $\sim$0.5-0.6\,keV. The equivalent  
blackbody emission radius of the hottest component is the smallest and 
that of the coolest component is the biggest. During dips, the 
fluxes of the two hot components are linearly correlated, while that of the third
component doesn't show any significant variation.
The Fe line flux remains constant within errors during the short dips. However, 
during the long dips the line flux changes significantly and is positively 
correlated with the fluxes of the two hot components. These results suggest: 
(1) the temperature of the X-ray emitting region decreases with radius, (2) the 
Fe K$\alpha$ line emitting region is close to the hot continuum emitting region,
and (3) the size of the Fe line emitting region is bigger than the size of the 
obscuring matters causing short dips but smaller than the
sizes of those causing long dips.
}

   \authorrunning{Y.-J. Lei et al.}            
   \titlerunning{X-ray Spectroscopy of the dips of Cir X-1}  

   \maketitle

%
%
\section{Introduction}           
\label{sect:intro}

The X-ray light curves of some low mass X-ray binaries (LMXBs)
contain dips, which usually occur  near phase zero. In some
cases, the dips may be due to the disk instability resulting in the
rapid removal and replenishment of matter forming the inner part of
an optically thick accretion disk (e.g., Belloni et al. 1997;
Greiner et al. 1996). However, in most cases, the dips are suggested
to be due to the absorption by the  matter passing through the line
of sight to the X-ray emitting region (e.g., White et al. 1995).
The X-ray spectral evolution during the
dips can provide us information about the geometry  and
physical conditions of these regions in LMXBs (e.g., Asai et al.
2000; Barnard et al. 2001).

The X-ray spectrum of a LMXB usually includes
 both the continuum and 6.4\,keV Fe K$\alpha$ line
emission components ({\sl e.g}., Hirano et al. 1987; White et al. 1986).
The continuum emission component is often described by the two
different models developed in the 1980's: the Western model
and   Eastern model. The ``Western model'' suggests that
Comptonization dominates the spectra, which has the form of a power
law with high energy cut off corresponding to the energy limit of
the Comptonizing electrons. However, for bright sources, an
additive  blackbody component representing the emission from the
neutron star (NS) is needed (e.g., White et al. 1988). The ``Eastern
model'' suggests that every observed spectrum contains two spectral
components: multi-temperature disk blackbody emission from the inner
disk and Comptonized emission provided by  the seed photons of NS
(e.g., Mitsuda et al. 1989). As declaimed, the physical conditions
of the emission region can be described by the ``Eastern model''.
For example, Done et al. (2002) assumed that there is an intrinsic
low energy cut off in the spectrum at $\sim$ 1\,keV due to
the lack of low energy seed photons, and the emission of NS  is
probably buried beneath an optically thick boundary layer.
As a result, the boundary layer dominates the hard  spectrum while
the disk dominates  the soft energies.
More recently, a model was proposed and developed by 
Church et al. (Church \& Ba\l uci\'nska-Church 1995; Church et al.
1997; Ba\l uci\'nska-Church et al. 1999),  in which two continuum
components exist in the dipping LMXBs: the  simple blackbody
emission from NS plus Comptonized emission from an extended
accretion disk corona (ADC) above the accretion disk and a
``progressive covering'' description of absorption. In addition,
this  model ascribes the dips to the absorption of obscuring matter
to the emission components.
For example, in X1755-338 and X1624-490, the absorption to the
blackbody component is primarily responsible for dipping (Church \&
 Ba\l uci\'nska-Church 1995, Church \&  Ba\l uci\'nska-Church
1996), while in X1658-298 the absorption to the blackbody component
is lower than that to the Comptonized component (Oosterbroek et al.
2001). In addition, for Cir X-1, Shirey et al. (1999) used a disk
blackbody  plus a blackbody and a ``progressive covering'' to fit the
spectra during the dips; Ding et al. (2006a) fitted the dip
spectra with the  two blackbodies plus a ``progressive covering'' for
a long dip of Cir X-1 (also see Brandt et al. 1996).

Fe K$\alpha$ line emission at 6.4 $\sim$ 6.7\,keV is a
common feature in the spectra of LMXBs, but its
origin is still in dispute. White et al. (1986) detected Fe K$\alpha$
emission line in five out of six LMXBs,
and proposed that the emission line is produced by the recombination
of Fe $XXVI$ in the inner ADC. Such a scenario was also suggested 
by Hirano et al. (1987). The Fe K$\alpha$ emission of dipping LMXBs 
is even more interesting since dips may provide new clues about
the location of the  line emitting region. 
Barnard et al. (2001) proposed that in XB 1323-619 the
Fe line at $\sim$ 6.4\,keV origins probably in the ADC,
while  Smale et al. (1993) ascribed the Fe line emission to the
reflection from the accretion disk itself for Cygnus X-2.
Using {\sl RXTE} data,  Shirey et al. (1999) investigated  the
spectral evolution of Cir X-1  during dips and found that the Fe
line flux is roughly constant during the dip transitions, suggesting
the Fe K$\alpha$ emission to be produced in the scattering medium.

Dips occur frequently on the light curve of Cir X-1, providing 
a good opportunity to study statistically the structure of the 
X-ray emitting regions and the location of Fe
K$\alpha$ line.
The spectral fits of the data of {\sl ASCA}, {\sl RXTE} and {\sl
BeppoSAX} during the dips are consistent with  a partial covering
model, in which the X-ray emission consists of a bright component
undergoing varying absorbing matter and a faint component unaffected
by the absorbing matter (e.g., Brandt et al 1996; Shirey et al. 1999; 
Iaria et al. 2001). The partial covering is a determinant of the behaviors
observed from Cir X-1, and matter
at the outer edge of the accretion disk, with an edge-on disk
orientation, could explain the partial covering of the spectrum (Brandt et al. 1996).
Ding et al. (2006a) studied a long dip of Cir X-1 and concluded
that the covering matter exists from the surface of NS to the emitting region
of iron line in the accretion disk.
However, another standpoints also exist, for example, D{\'{\i}}az Trigo et al. (2006) 
showed that there is no need for partial
covering to explain the spectral evolution during the dips with {\sl XMM} data.

In those previous studies of the dips of Cir X-1, each paper is usually based on 
the spectral analysis of a particular dip. We therefore present herein  a 
statistical study of the spectral evolution of Cir X-1 during various dips so as 
to better constrain the structure of the X-ray
emitting regions. The paper  is organized as follows: the observations, spectral models, 
and data analyses are described in Sect. 2, the results are shown in Sect. 3, 
the discussions are preseted in
Sect. 4 and the summary is in Sect. 5.

\section{Observations and Analysis}
\label{sect:Obs}
\subsection{Observations and Light Curves}
\label{subsect:data}
The Proportional Counter Array (PCA) of {\sl RXTE} consists of 5 non-imaging,
coaligned Xe multiwire proportional counter units (PCUs). It has a field of view
of 1$^{\circ}$ $\times$ 1$^{\circ}$ and a total
collecting area of $\sim$ 6500\,cm$^{2}$.
We use five datasets collected by PCU0 and PCU2 when both of them are working,
and both long (duration $\geq$ 10$^{4}$\,s) and short (duration $\leq$ 10$^{4}$\,s) 
dips are detected  in these data (Table 1). 
Since we are interested in the spectral evolution during dips, we
only study the  dips with durations $\geq$ 10$^{2}$\,s so
that each dip can be divided into a few segments and each segment
contains enough counts for spectral analysis.
Altogether we study the spectral properties of three long dips and
two short dips that are divided each into 5-17 segments.
Two typical light curves  with the denoted segments are shown in
Figs. 1 and 2.

The data are
reduced and analyzed using the {\sl Ftools} software package version
v5.2. We filter the data using the standard {\sl RXTE} criteria,
selecting time intervals for which parameters ${\tt Elevation\_Angles}
<10\degr$, ${\tt Time\_Since\_SAA} \geq 30$ min, ${\tt
Pointing\_Offsets} < 0.02 \degr$, and the background 
electron rate ${\tt Electron2} < 0.1$. 
Events in the {\sl RXTE} energy range $\sim$ 2.5-25\,keV are selected for
the spectral analysis. PCA background subtraction is carried out using the latest
versions of the appropriate background models, and a 1\% systematic error
is added to the spectra to account for calibration uncertainties.
As usual, the spectral fitting software $XSPEC$ is used.

\begin{table*}
\caption[]{Selected RXTE observations of Cir X--1\label{tbl-1}}
\begin{tabular}{l l l l}
\hline \hline
   OBSID              &  Start Time  &    Stop  Time   &     Exposure(s)\\
\hline
  10122-01-04-00      & 96-9-20 07:47:01 & 96-9-20 12:30:13 &    16992  \\
\hline
  10122-01-04-02      & 96-9-21 04:40:02 & 96-9-21 11:01:13 &    22871  \\
\hline
  30080-01-01-000     & 98-03-03 16:42:14 & 98-03-03 23:22:07 &    23993  \\
\hline
  30081-06-01-000     & 98-10-05 00:07:41 & 98-10-05 08:01:07 &    28406  \\
  30081-06-01-01      & 98-10-04 22:31:54 & 98-10-04 23:36:14 &    3860   \\
\hline
  60024-01-01-00      & 01-02-28 06:54:03 & 01-02-28 09:10:15 &    8172   \\
  60024-01-01-01      & 01-02-28 03:40:03 & 01-02-28 05:59:15 &    8352   \\
  60024-01-01-02      & 01-02-28 02:00:41 & 01-02-28 02:48:15 &    2854   \\
  60024-02-01-03      & 01-02-28 09:41:03 & 01-02-28 10:10:15 &    1752   \\
\hline
\end{tabular}
\end{table*}

\begin{figure*}
\begin{center}
\includegraphics[width=6cm,angle=270,clip]{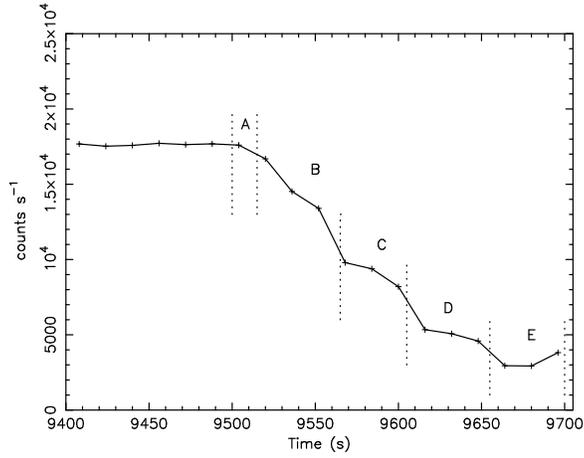}
\caption{Typical light curve (16\,s bins) of short ($<10^{4}\,s$) dip 
during dip and dip
transition from PCA observations of Cir X-1 (OBSID:30080). For clarity,
the errors are not plotted.}
\label{fig1}
\end{center}
\end{figure*}

\begin{figure}
\begin{center}
\includegraphics[width=6cm,angle=270,clip]{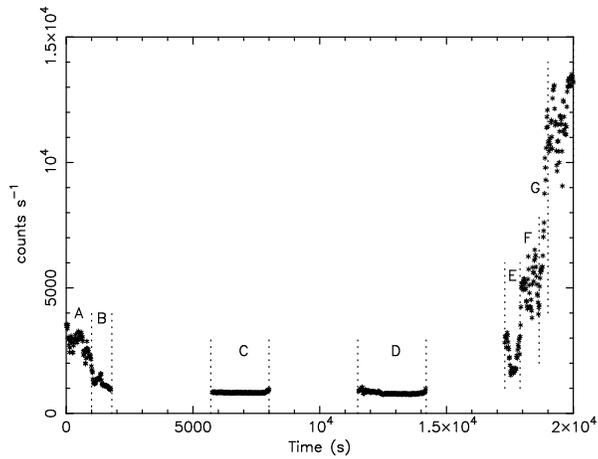}
\caption{Typical light curve (16\,s bins) of long ($>10^{4}\,s$) dip during dip and dip
transition from PCA observations of Cir X-1 (OBSID:30081).}
\label{fig2}
\end{center}
\end{figure}

\subsection{Spectral Model}
\label{subsect: model}
In the literature various spectral models have been used to fit the
spectra of Cir X-1 during dips (e.g., Brandt et al. 1996; Shirey et al. 1999;
Iaria et al. 2002; Ding et al. 2006a).
Brandt et al. (1996) used a model containing two blackbody
components and a partial covering. The partial covering can affect the hot
component, two or none of them.
Shirey et al. (1999) used a model which can be expressed as:
$F=[exp(-\sigma_{ph}N_{H}^{(1)})exp(-\sigma_{Th}N_{H}^{(1)})+exp(-\sigma_{ph}N_{H}^{(2)})f]M$,
where $F$ is the observed flux, $\sigma_{ph}$ and $\sigma_{Th}$ are
the photoelectric and Thomson cross sections, $N_{H}^{(1)}$ and
$N_{H}^{(2)}$ are the effective absorption hydrogen column densities to the
bright and faint components, respectively, $f$ is the ratio of the
unabsorbed flux of the faint component to the unabsorbed flux of the
bright component, and $M$ is the disk blackbody plus blackbody
model. Ding et al. (2006a) fitted the dip spectra with 
the ``Eastern'' model and a modified ``Western''
model. 

However, we find that all these above mentioned spectral models 
are improper to reveal the physical conditions of the dips. 
In Brandt et al. (1996) and Ding et al. (2006a), one of
 the fitted temperatures changes greatly within a dip. For an example, 
the temperature of the disk-blackbody  component
changes suddenly from $\sim$0.6\,keV to $\sim$1.2\,keV when its detected
fluxes evolves from the dip bottom to the transition state.  Such 
temperature changes seem unlikely if the dips are mainly due to the 
obscuration as widely accepted, because the obscuring matter can not
change the physical state of the emission region. 
In the model of Shirey et al. (1999), the residual flux at the bottom
of the dips is a fraction of the total emission. However, the 
hardness-intensity diagram (Fig. 3 in their paper and this paper)
shows that the detected spectrum of Cir X-1 at the lowest flux level is
even softer than that of its out-of-dip. This means that the residual 
emission spectrum at dip bottom is intrinsically softer  than that
nondip spectrum, i.e., it is improper to assume that
the residual emission is simply a small fraction of the total 
emission.

The soft spectrum at the bottom of the dip and the sudden temperature 
increase of the disk-blackbody component (Ding et al. 2006a) when Cir 
X-1 moves to the transition state, imply that there exists a weak and
soft component which is not affected by the obscuring matter. When
there is no dip or in the transition state, the usually called 
``NS blackbody emission and disk-blackbody emission'' dominate, and so 
the weak component is difficult to be noted. When most of the ``NS blackbody emission 
and disk-blackbody emission'' are obscured when Cir X-1 is at the 
bottom of the dips, the contribution of the weak and soft component
to the total flux becomes significant or even dominant, and so the
detected spectrum at the dip bottom is softer than the
out-of-dip spectrum. Such a weak and soft component
is also noted by Ding et al. (2006a) but they do not consider it further
in the spectral fitting process.

Taking the above facts into account, we use in this paper a spectral
model containing three blackbody components and a Gaussian at 
the energy of Fe K$\alpha$ line. The 
first two blackbody components with partial covering are similar to those models 
in Brandt et al. (1996),  Shirey et al. (1999) and Ding et al. (2006a), 
while the additional third component represents the weak and soft emission that are
free of partial covering absorption. The expression of our model is: 
$F=wabs[pcfabs*cabs(bbodyrad(1)+bbodyrad(2)+gaussian)+bbodyrad(3)]$,
where $wabs$ represents the photoelectric absorption of the
interstellar medium (ISM), $pcfabs$ corresponds to the partial
covering photoelectric absorption, $cabs$ accounts for the Compton
scattering of the partial covering matter,
$bbodyrad(1)$ and $bbodyrad(2)$ represent the
continuum emission which is affected by the partial covering, $bbodyrad(3)$
is responsible for the soft faint emission component, and $gaussian$ accounts
for the Fe K$\alpha$ line emission.
As shown later, the fitted parameters by this model are
much more reasonable than those in Brandt et al. (1996), Shirey et al. (1999) 
and Ding et al. (2006a).

In the spectral fitting, the column density of $wabs$ is fixed as
1.8$ \times 10^{22}$\,cm$^{-2}$ as determined in the
previous work (e.g., Brandt et al. 1996; Iaria et al. 2001), and the
column densities of $pcfabs$ and $cabs$ are forced to be identical (Shirey et al. 1999). 
As shown by the following fitting results (Figs. 4-9, Tables 2-3), such a model gives 
statistically acceptable fits to both the dip and nondip spectra,  and the
obtained parameters look physically reasonable.

%
\begin{figure}
\begin{center}
\includegraphics[width=6cm,angle=270,clip]{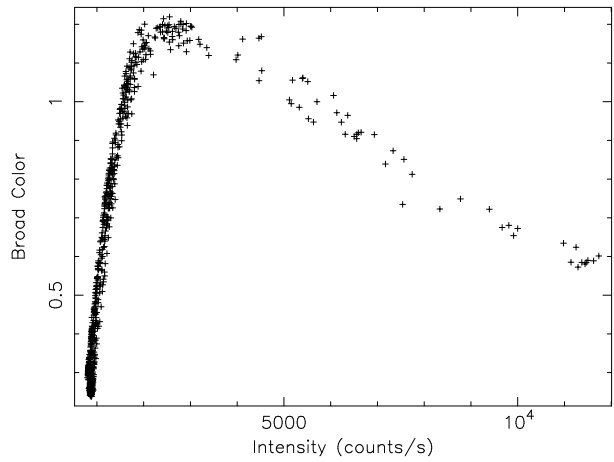}
\caption{Hardness-intensity diagram for data in Fig. 2, intensity is I(2.0-18\,keV),
the hardness ratio is defined as : I(6.3-18\,keV)/I(2.0-6.3\,keV). 
Each point represents 16\,s of
background-subtracted data from all five PCA detectors (also see, Shirey et al. 1999).}
\label{fig3}
\end{center}
\end{figure}

\begin{figure}
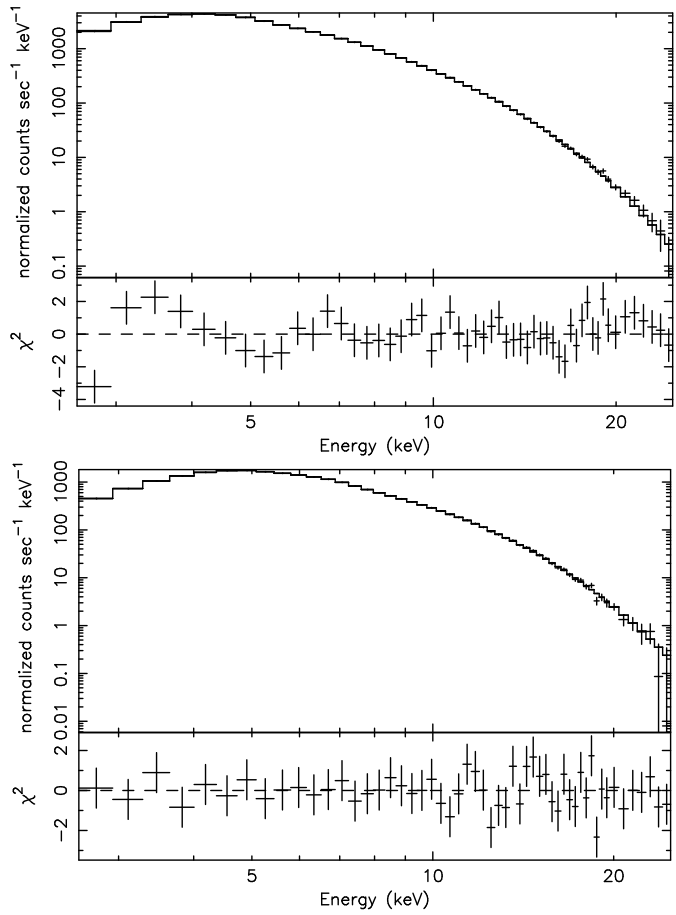

\begin{center}
\includegraphics[width=6cm,angle=270,clip]{f4a.ps}
\includegraphics[width=6cm,angle=270,clip]{f4b.ps}
\caption{Top: Typical spectral fit for the dip spectra (spectrum C of OBSID 30080 in Table 2).
Bottom: Typical spectral fit for the nondip spectra (nondip data of OBSID 30080),
with $\chi^2$ values of 55.4 for 54 degrees of freedom.}
\label{fig4}
\end{center}
\end{figure}

\section{Result}
\label{sect:Result}
The spectra of all the segments have been fitted with the model
described in Sect. 2. Tables 2 and 3 list the
spectral fitting results for segments in OBSIDs 30080 and
30081. In the following we will present our results in various aspects.

\begin{table*}
\caption[]{The Fitted Parameters for Spectra from the OBSID 30080 dip}
\begin{tabular}{lccccc}
\hline \hline
Fit Parameters & A & B & C & D & E  \\
\hline
$N_{H}^{a}$ &
           1.8(fixed)              &     1.8(fixed)       &
       1.8(fixed)              &     1.8(fixed)       &
       1.8(fixed)               \\
\hline
$N_{H}^{b}$ &
    $0.0_{-0.0}^{+6.3}$     & $2.4_{-0.3}^{+0.6}$  &
    $16_{-8}^{+3}$          & $34_{-5}^{+6}$       &
    $38_{-21}^{+13}$           \\
\hline
$Percentage^c$ &
    $4.85_{-4.85}^{+95.15}$      &   95.2(fixed)            &
    $75.0_{-4.0}^{+24.4}$      & $89.9_{-2.9}^{+2.8}$     &
    $81.0_{-5.1}^{+4.2}$        \\
\hline
$kT_{\rm 1}^d$ (keV) &
    $1.85_{-0.04}^{+0.04}$     & $1.87_{-0.05}^{+0.05}$   &
        $1.91_{-0.08}^{+0.07}$   &
    $1.97_{-0.08}^{+0.10}$     &
    $1.89_{-0.08}^{+0.08}$      \\
$A_{\rm 1}^e$ &
    $107_{-16}^{+9}$              & $93.7_{-8.8}^{+22.3}$    &
    $80.1_{-22.7}^{+21.4}$        & $61.9_{-10.7}^{+23.3}$     &
    $81.7_{-23.2}^{+36.3}$        \\
\hline
$flux_{\rm 1}^m$ &
    $1.11_{-0.16}^{+0.22}$       & $0.85_{-0.15}^{+0.22}$    &
    $0.49_{-0.13}^{+0.21}$        & $0.21_{-10.7}^{+23.3}$     &
    $0.26_{-0.08}^{+0.12}$        \\

\hline
$E_{\rm line}^f$ (keV) &
           6.60(fixed)         & $6.56_{-0.28}^{+0.33}$     &
    $6.48_{-0.49}^{+0.47}$     &
           6.60(fixed)     &
    $6.64_{-0.26}^{+0.34}$              \\
\hline
$\sigma^g$ (keV) &
          0.5(fixed)           &      0.5(fixed)       &
    $0.58_{-0.58}^{+0.70}$     &
    $0.36_{-0.36}^{+0.21}$     &   $0.22_{-0.22}^{+0.40}$     \\
\hline
$flux^h$ &
        $12.93_{-8.71}^{+5.76}$     &
    $16.98_{-9.21}^{+4.95}$    &
    $22.33_{-14.39}^{+47.80}$     & $14.14_{-2.95}^{+8.10}$ &
    $14.65_{-4.29}^{+16.54}$               \\
\hline
$kT_{\rm 2}^i$ (keV) &
    $1.00_{-0.04}^{+0.03}$     &
        $1.03_{-0.04}^{+0.01}$     &
    $1.06_{-0.12}^{+0.05}$     &
    $1.12_{-0.05}^{+0.06}$     &
    $1.08_{-0.04}^{+0.06}$      \\
$A_{\rm 2}^j$ &
    $ 3.36_{-0.33}^{+0.43}\times10^{3} $     & $ 2.74_{-0.30}^{+0.25}\times10^{3} $   &
    $ 2.19_{-0.41}^{+0.58}\times10^{3} $      & $ 1.69_{-0.21}^{+0.42}\times10^{3} $   &
    $ 1.65_{-0.81}^{+0.95}\times10^{3} $        \\

\hline
$flux_{\rm 2}^n$ &
    $3.92_{-0.33}^{+0.43} $        & $2.65_{-0.28}^{+0.48}$   &
    $1.35_{-0.51}^{+0.50}$        & $0.57_{-0.14}^{+0.17}$    &
    $0.57_{-0.28}^{+0.34}$             \\
\hline
$kT_{\rm 3}^k$ (keV) &
          0.50(fixed)    &        0.50(fixed)    &
          0.50(fixed)          & $0.49_{-0.02}^{+0.02}$   &
          0.50(fixed)       \\
$A_{\rm 3}^l$ &
    $4.60_{-4.38}^{+8.38}\times10^{3}$        & $0.0_{-0.0}^{+2.93}\times10^{3}$   &
    $0.02_{-0.02}^{+6.47}\times10^{3}$        & $8.45_{-3.83}^{+7.08}\times10^{3}$    &
    $5.32_{-2.16}^{+3.87}\times10^{3}$             \\
\hline
$flux_{\rm 3}^p$ &
    $0.19_{-0.18}^{+0.35} $        & $0.0_{-0.0}^{+0.12}$   &
    $0.0_{-0.0}^{+0.28}$        & $0.31_{-0.14}^{+0.26}$    &
    $0.22_{-0.07}^{+0.13}$             \\
\hline
$\chi^2~(dof)^p$ &
    68.63(46)                  & 42.82(46)               &
    35.51(44)                  & 30.46(45)               &
    24.04(44)                   \\
\hline
$\chi_{\nu}^{2q}$ &
    1.49                       & 0.93                     &
    0.81                       & 0.68                     &
    0.55                       \\
\hline
\end{tabular}
\vspace{0.2cm}
\parbox{6.8in}
{\baselineskip 5pt {\sc Note~} Model of wabs[pcfabs*cabs(bbodyrad(1)+gaussian+bbodyrad(2))]+wabs*bbodyrad(3) is used.
Uncertainties are given at 90\% confidence level.\\}

$^a$Interstellar column density, in units of $10^{22}$\,cm$^{-2}$\\
$^b$Partial covering column density, in units of $10^{22}$\,cm$^{-2}$\\
$^c$Partial covering percentage\\
$^d$The temperature of blackbody(1)\\
$^e$The normalization of blackbody(1)\\
$^m$The flux of blackbody(1) in the 2.0-10.0\,keV energy range and in units of photons
cm$^{-2}$\,s$^{-1}$\\
$^f$Gaussian line energy\\
$^g$Gaussian width\\
$^h$Flux of the gaussian line, in units of $10^{-3}$ photons\,cm$^{-2}$\,s$^{-1}$\\
$^i$The temperature of blackbody(2)\\
$^j$The normalization of blackbody(2)\\
$^n$The flux of blackbody(2) in the 2.0-10.0\,keV energy range and in units of photons\,
cm$^{-2}$\,s$^{-1}$\\
$^k$The temperature of blackbody(3)\\
$^l$The normalization of blackbody(3)\\
$^p$The flux of blackbody(3) in the 2.0-10.0\,keV energy range and in units of photons\,cm$^{-2}$\,s$^{-1}$\\
$^p$$\chi^2~(dof)$: $\chi^2$: Chi-Squared; $dof$: degrees of freedom\\
$^q$$\chi_{\nu}^2$: reduced chi-squared: $\chi^2$/$dof$\\
\end{table*}

\begin{landscape}
\begin{table*}
\caption[]{The Fitted Parameters for Spectra of the OBSID 30081 dip}
\begin{tabular}{lccccccc}
\hline \hline
Fit Parameters & A & B & C & D & E & F & G  \\
\hline
$ N_{H}^{a}$ &
       1.8(fixed)              &     1.8(fixed)       &
       1.8(fixed)              &     1.8(fixed)       &
       1.8(fixed)               &
       1.8(fixed)              &
       1.8(fixed)               \\
\hline
$ N_{H}^{b}$ &
    $43_{-4}^{+3}$             & $81_{-42}^{+29}$    &
    $325_{-53}^{+39}$          & $247_{-40}^{+29}$    &
    $44_{-8}^{+10}$             &
    $23_{-4}^{+5}$             &
    $13_{-8}^{+5}$             \\
\hline
$Percentage^c$ &
    $98.1_{-9.2}^{+1.9}$       & $82.9_{-7.1}^{+7.6}$     &
    $89.6_{-5.3}^{+2.8}$       & $90.8_{-5.4}^{+2.8}$     &
    $98.0_{-7.3}^{+2.0}$       &
         95.2(fixed)           &
    $55.1_{-5.5}^{+44.9}$                \\
\hline
$kT_{\rm 1}^d$ (keV) &
    $2.14_{-0.03}^{+0.04}$     & $2.11_{-0.07}^{+0.08}$   &
     2.20(fixed)               & $2.25_{-0.19}^{+0.42}$   &
    $2.11_{-0.05}^{+0.05}$     &
    $2.10_{-0.04}^{+0.04}$     &
    $2.05_{-0.04}^{+0.03}$     \\
$A_{\rm 1}^e$ &
    $37.3_{-5.1}^{+5.0}$       & $31.0_{-13.9}^{+45.4}$    &
    $36.9_{-5.6}^{+8.7}$       & $33.5_{-24.8}^{+29.9}$     &
    $37.7_{-7.4}^{+9.7}$       &
    $41.6_{-5.8}^{+6.8}$       &
    $50.0_{-9.3}^{+4.2}$        \\
\hline
$flux_{\rm 1}^m$ &
    $0.11_{-0.02}^{+0.02}$       & $0.06_{-0.03}^{+0.09}$    &
    $0.00_{-0.00}^{+0.00}$       & $0.01_{-0.01}^{+0.01}$     &
    $0.10_{-0.02}^{+0.32}$       &
    $0.23_{-0.03}^{+0.04}$       &
    $0.47_{-0.09}^{+0.04}$        \\
\hline
$E_{\rm line}^f$ (keV) &
     6.60(fixed)               &   $6.63_{-0.13}^{+0.11}$   &
    $6.56_{-0.04}^{+0.06}$     &
    $6.55_{-0.05}^{+0.05}$     &
    $6.64_{-0.20}^{+0.32}$     &
    $6.65_{-0.24}^{+0.27}$     &
    $6.42_{-0.40}^{+0.32}$              \\
\hline
$\sigma^g$ (keV) &
     0.50(fixed)               &   $0.38_{-0.13}^{+0.18}$     &
    $0.20_{-0.20}^{+0.12}$     &
    $0.26_{-0.15}^{+0.09}$     & $0.45_{-0.31}^{+0.24}$   &
    $0.39_{-0.39}^{+0.28}$     &
    $0.54_{-0.43}^{+0.30}$          \\
\hline
$flux^h$ &
    $7.36_{-3.40}^{+2.88}$     & $8.17_{-2.60}^{+4.86}$     &
    $4.90_{-2.81}^{+2.54}$     &
    $5.64_{-2.77}^{+3.94}$     & $14.63_{-6.14}^{+3.74}$ &
    $17.21_{-7.34}^{+12.25}$   &
    $29.31_{-13.68}^{+16.96}$                \\
\hline
$kT_{\rm 2}^i$ (keV) &
    $1.19_{-0.03}^{+0.03}$     & $0.84_{-0.22}^{+0.13}$           &
    $1.27_{-0.07}^{+0.07}$     &
        $1.29_{-0.08}^{+0.13}$     &
    $1.21_{-0.05}^{+0.06}$     &
    $1.17_{-0.03}^{+0.03}$     &
    $1.12_{-0.03}^{+0.02}$     \\
$A_{\rm 2}^j$ &
    $ 0.89_{-0.15}^{+0.19}\times10^{3}$         & $ 2.13_{-1.35}^{+4.86}\times10^{3} $   &
    $ 4.11_{-1.73}^{+3.31}\times10^{3}$        & $ 2.15_{-1.09}^{+1.39}\times10^{3} $   &
    $ 0.68_{-0.17}^{+0.50}\times10^{3} $        &
    $ 1.16_{-0.26}^{+0.30}\times10^{3} $        &
    $ 1.33_{-0.46}^{+0.34}\times10^{3} $        \\
\hline
$flux_{\rm 2}^n$ (keV) &
    $0.23_{-0.04}^{+0.05}$     & $0.12_{-0.07}^{+0.26}$          &
    $0.09_{-0.05}^{+0.07}$     &
        $0.09_{-0.04}^{+0.05}$     &
    $0.17_{-0.08}^{+0.13}$     &
    $0.61_{-0.16}^{+0.16}$     &
    $1.40_{-0.47}^{+0.35}$      \\
\hline
$kT_{\rm 3}^k$ (keV) &
    $0.58_{-0.02}^{+0.02}$     & $0.54_{-0.07}^{+0.07}$   &
    $0.53_{-0.02}^{+0.01}$     & $0.51_{-0.01}^{+0.03}$   &
    $0.54_{-0.02}^{+0.02}$     &
    $0.48_{-0.09}^{+0.11}$     &
          0.50(fixed)              \\
$A_{\rm 3}^l$ &
    $3.61_{-0.62}^{+0.70}\times10^{3}$        & $4.11_{-1.57}^{+2.51}\times10^{3}$   &
    $4.71_{-0.71}^{+1.06}\times10^{3}$        & $5.33_{-0.54}^{+0.97}\times10^{3}$   &
    $4.32_{-0.91}^{+1.23}\times10^{3}$        &
    $6.75_{-4.44}^{+18.16}\times10^{3}$       &
    $0.0_{-0.0}^{+6.85}\times10^{3}$          \\
\hline
$flux_{\rm 3}^p$ &
    $0.35_{-0.06}^{+0.06}$        & $0.26_{-0.08}^{+0.15}$    &
    $0.28_{-0.04}^{+0.03}$        & $0.26_{-0.05}^{+0.05}$    &
    $0.28_{-0.06}^{+0.08}$        &
    $0.22_{-0.16}^{+0.60}$        &
    $0.0_{-0.0}^{+0.23}$          \\
\hline
$\chi^2~(dof)^p$ &
    17.68(45)                  & 20.32(43)               &
    23.32(44)                  & 21.17(43)               &
    15.54(43)                  &
    18.28(44)                  &
    18.45(44)                   \\
\hline
$\chi_{\nu}^{2q}$ &
    0.39                       & 0.47                     &
    0.53                       & 0.49                     &
    0.36                       &
    0.42                       &
    0.42                        \\
\hline

\end{tabular}
\vspace{0.2cm}
\parbox{6.8in}
{\baselineskip 5pt {\sc Note~} Model of wabs[pcfabs*cabs(bbodyrad(1)+gaussian+bbodyrad(2))]+wabs*bbodyrad(3) is used. Uncertainties
are given at 90\% confidence level for the derived parameters of the model applied.\\}

$^a$Interstellar column density, in units of $10^{22}$\,cm$^{-2}$\\
$^b$Partial covering column density, in units of $10^{22}$\,cm$^{-2}$\\
$^c$Partial covering percentage\\
$^d$The temperature of blackbody(1)\\
$^e$The normalization of blackbody(1)\\
$^m$The flux of blackbody(1) in the 2.0-10.0\,keV energy range and and in units of photons\,
cm$^{-2}$\,s$^{-1}$\\
$^f$Gaussian line energy\\
$^g$Gaussian width\\
$^h$Flux of the gaussian line, in units of $10^{-3}$ photons\,cm$^{-2}$\,s$^{-1}$\\
$^i$The temperature of blackbody(2)\\
$^j$The normalization of blackbody(2)\\
$^n$The flux of blackbody(2) in the 2.0-10.0\,keV energy range and in units of photons\,
cm$^{-2}$\,s$^{-1}$\\
$^k$The temperature of blackbody(3)\\
$^l$The normalization of blackbody(3)\\
$^p$The flux of blackbody(3) in the 2.0-10.0\,keV energy range and in units of photons
\,cm$^{-2}$\,s$^{-1}$\\
$^p$$\chi^2~(dof)$: $\chi^2$: Chi-Squared; $dof$: degrees of freedom\\
$^q$$\chi_{\nu}^2$: reduced chi-squared: $\chi^2$/$dof$\\
\end{table*}
\end{landscape}

\subsection{Partial covering column density and the 
correlations with other  parameters}
\label{subsect: partial covering}
Brandt et al. (1996) used a partial covering model to fit the spectra
before and after the dip transition of Cir X-1, and found a strong Fe
K$\alpha$ edge in the low-state spectra,
which indicates that the obscuring matter has a very high column density,
i.e, $N_{\rm H}$ $>$ 10$^{24}$\,cm$^{-2}$. Shirey et al. (1999) obtained 
a  similar result. Our results show that the hydrogen column density of 
partial covering absorption is between 10$^{22}$\,cm$^{-2}$ and  
3.3 $\times$ 10$^{24}$\,cm$^{-2}$ (e.g., Fig. 5).
The large variation of column density could be due to the partial
covering and/or the inhomogeneity of the covering matter. 

Figures 5 and 6 show relations of the normalizations of 
the Fe K$\alpha$, blackbody(1) and blackbody(2) versus the 
partial covering absorption column density. 
Obviously, the normalization of Fe K$\alpha$ line has a positive
correlation with $N_{\rm H}$, while the correlations are weak
for the normalizations of blackbody(1) and blackbody(2). Since in
the spectral fitting, the absorption column density and normalization
are usually coupled with each other, in order to check whether
the correlation between the Fe K$\alpha$ strength and the absorption 
column density is real or not, we derive the normalization ratios 
of the Fe K$\alpha$ normalization to the normalizations of blackbody(1)
and blackbody(2). As shown in Figs. 7 and 8, such coupling-free ratios
are also correlated with the absorption column density, indicating
that the origin of the Fe emission line is related to the
obscuration matter.

\begin{figure}
\begin{center}
\includegraphics[width=10cm,clip]{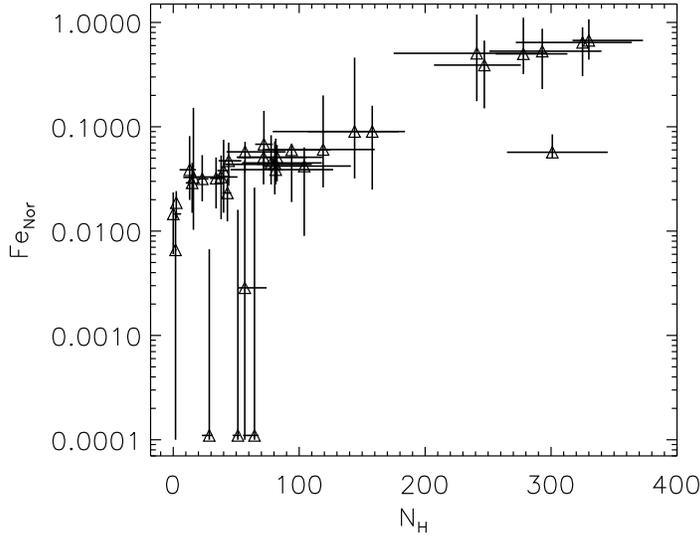}
\caption{The
normalization of Fe line is plotted against $N_{\rm H}$ (in units of
$10^{22}$\,cm$^{-2}$), and an obvious correlation can be found.}
\label{fig5}
\end{center}
\end{figure}
\begin{figure}
\begin{center}
\includegraphics[width=10cm,clip]{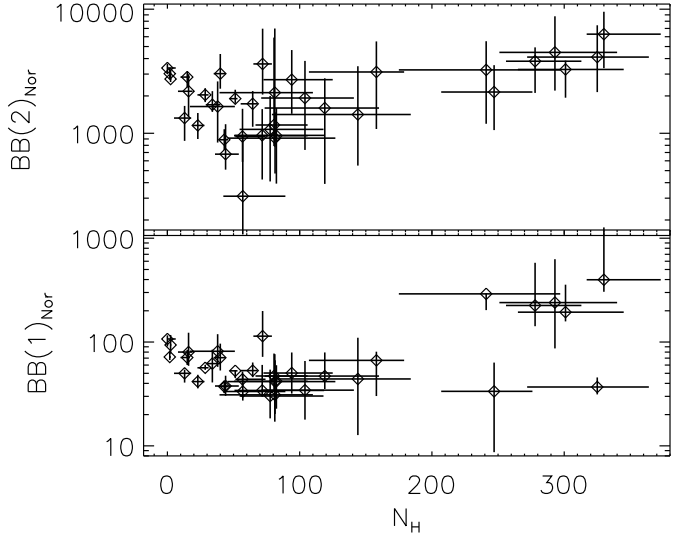}
\caption{The
normalizations of blackbody(1) (bottom) and blackbody(2) (top) are
plotted against $N_{\rm H}$, and both show the correlations.}
\label{fig6}
\end{center}
\end{figure}
\begin{figure}
\begin{center}
\includegraphics[width=10cm,clip]{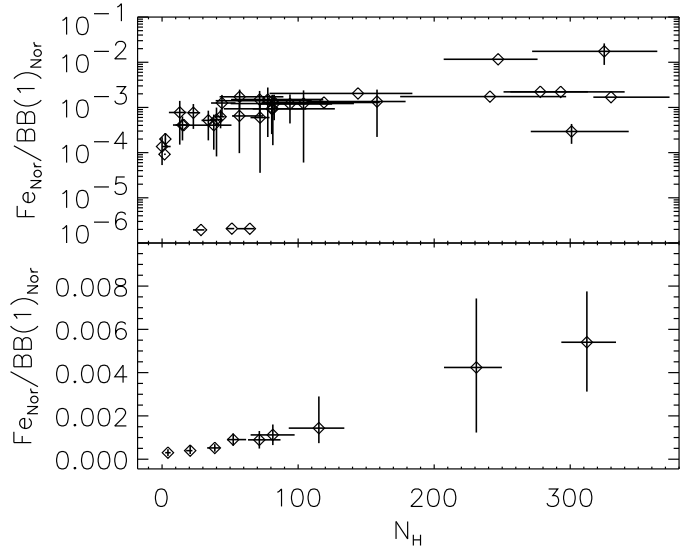}
\caption{Ratio of the normalization of Fe line to that of
blackbody(1) is plotted against $N_{\rm H}$. The bottom panel is a
four-point binned plot of the top one, a positive
correlation can be seen.}
\label{fig7}
\end{center}
\end{figure}

\begin{figure}
\begin{center}
\includegraphics[width=10cm,clip]{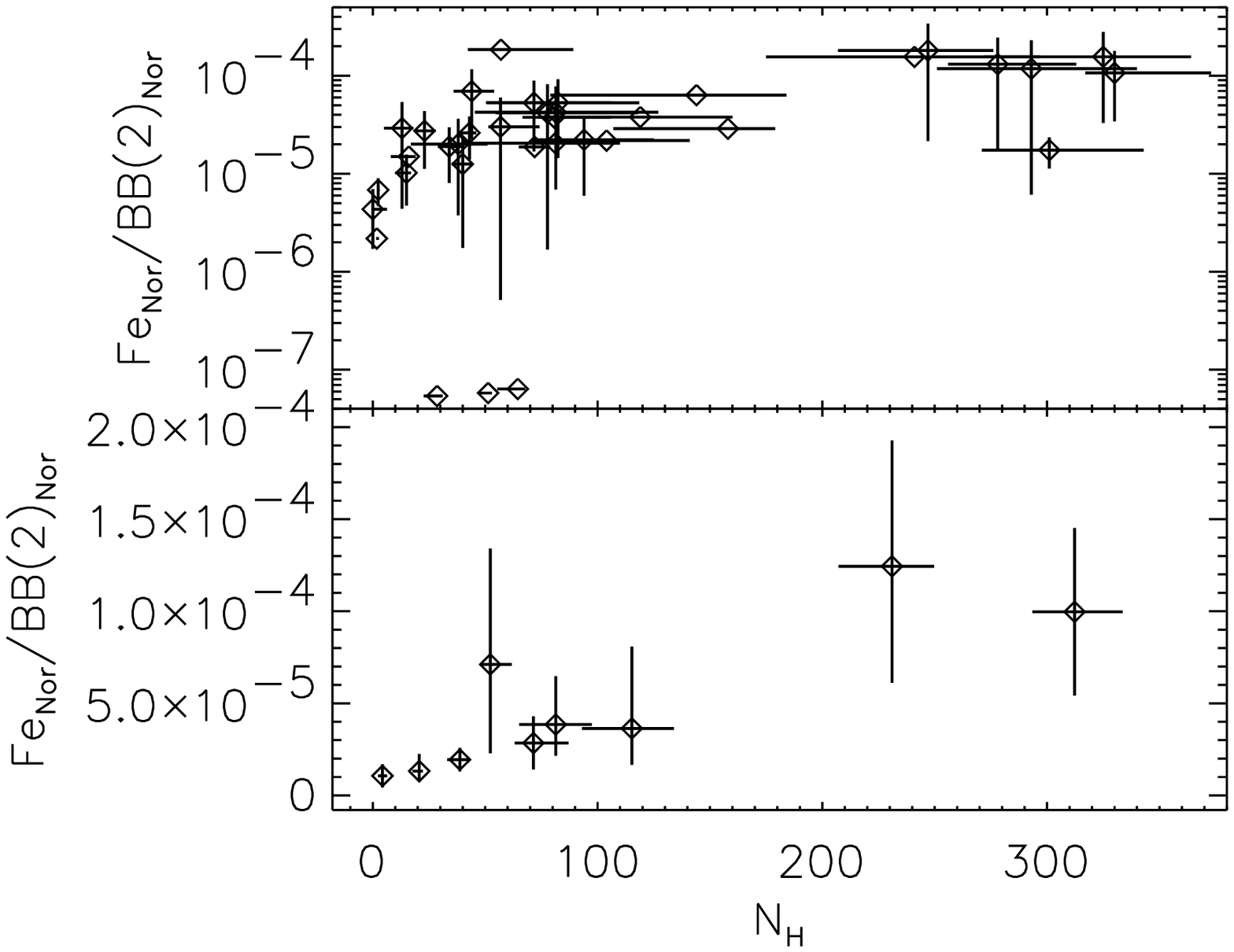}
\caption{Similar to Fig. 7, but  blackbody(2)
is substituted.}
\label{fig8}
\end{center}
\end{figure}

\subsection{Temperatures and Sizes of emission regions}
\label{subsect: temperatures and sizes}
For every dip, the temperature of blackbody(1) is about 2\,keV and that
of blackbody(2) is about 1\,keV, which is consistent with the  results of
Shirey et al. (1999) with {\sl RXTE}. 
The temperature of blackbody(3), unaffected by the partial covering,
is about 0.5$\sim$0.6\,keV, which is consistent with
the result of Brandt et al. (1996) with {\sl ASCA}.
The almost constant values of the temperature also show that the selecting 
of a three-blackbody-component spectral model is proper. 

The distance to Cir X-1 is about 6-10\,kpc (e.g., Stewart et al. 1991; 
Goss \& Mebold 1977), so we adopt a value of 8 kpc in converting the blackbody
normalization  to radius.
Hence, the obtained radii of blackbody(1) (blackbody(2)) are
mostly in the range of 4-7\,km (16-30\,km).
As for the blackbody(3), usually, its radii lie in the range of
30-50\,km. Since the typical radius of a NS is 10\,km, blackbody(1) may come 
from hot regions near/on the NS surface, while blackbody(2) and blackbody(3) 
are perhaps from the inner and outer disk respectively.

\subsection{Flux and Flux correlations}
\label{subsect: correlation}
In Fig. 9 we plot the relations between the fluxes of blackbody(1),
blackbody(2) and blackbody(3). A strong correlation between the 
fluxes of blackbody(1) and blackbody(2) is found, with a correlation
coefficient $0.98_{-0.04}^{+0.02}$. 
However, further investigations have not shown any sign of 
clear correlations between the flux of blackbody(3) and that of
blackbody(1) or blackbody(2).

We now show that the linear correlation between the fluxes
of blackbody(1) and blackbody(2) is intrinsic other than a false 
appearance of the spectral fitting. In the spectral fitting, 
blackbody(1) and 
blackbody(2) are all coupled with the partial covering. When the 
partial covering column 
density (and/or covering fraction) is somehow overestimated, 
the normalizations of the two components will be in turn higher than
the real values, and vice versa. Then the two normalizations
will show an artificial positive correlation. However, the fluxes plotted in
Fig. 9 are the absorbed ones (corresponding to the observed fluxes), 
in which the model coupling effect is eliminated. Therefore, the
linear correlations shown in Fig. 9 is reliable. 

The flux correlation between blackbody(1) and blackbody(2) implies
that these two components are usually obscured simultaneously. 
So their emission regions are probably very close to each other.
This is well consistent with the conclusions we drew from the equivalent
emitting radii in the previous section.

\begin{figure}
\begin{center}
\includegraphics[width=10cm,clip]{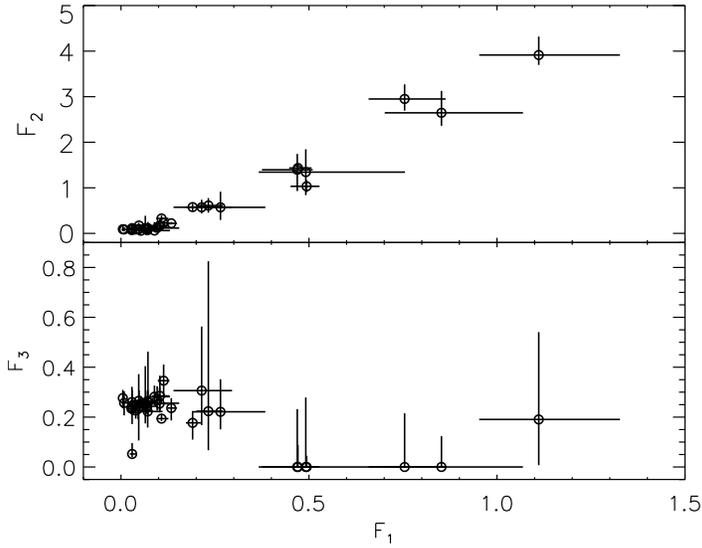}
\caption{The flux
(in units of photons\,cm$^{-2}$\,s$^{-1}$ ) of blackbody(1)
plotted against that of blackbody(2) (top panel) and blackbody(3)
(bottom panel). The top panel clearly shows a linear correlation,
whereas  no significant relationship has been revealed in the bottom
panel.}
\label{fig9}
\end{center}
\end{figure}

\subsection{Fe line flux and its correlations with continuum components}
\label{subsect: Fe line flux}
Figure 10 plots the Fe line fluxes versus the fluxes of the 
continuum components. It is found that the flux of the Fe line 
has a trend to increase along with the fluxes of blackbody(1) 
and blackbody(2), while no clear correlation
exists between the Fe line flux and that of blackbody(3).
Since the flux changes for both the line and continuum 
emission are mainly due to the obscuration of partial covering matters,
this correlation suggests that the Fe line emission origins close
to the hot regions that emit blackbody(1) and blackbody(2). 

We further divide the dips into long and short dips, and for
each category, we investigate the above correlations. As plotted in
Fig. 11, for long dips, the flux of Fe line has now even better 
correlations with those of the blackbody(1)
and blackbody(2) than the correlations for all the dips. However,
for short dips, such correlations do not exist anymore (see Fig. 12),
which is consistent with the results of Shirey et al. (1999). 
Similar to that for all the dips, correlations between the flux of the Fe line
and that of blackbody(3) are not detected either for long dips or for
short dips. These results tell us
 that the emitting region of the Fe line is smaller than the the obscuration matters
of long dips but bigger than those of the short dips.

\begin{figure*}
\begin{center}
\includegraphics[width=13cm,clip]{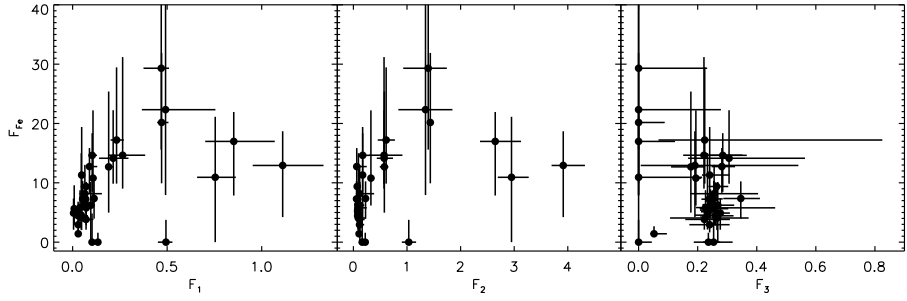}
\caption{Fe line
flux (in units of 10$^{-3}$ photons\,cm$^{-2}$\,s$^{-1}$)
plotted as a function of the flux of  blackbody(1) (left),
blackbody(2) (middle) and blackbody(3) (right),  for  all chosen
dips (10122-dip1, 30081, 60024, 10122-dip2, 30080).}
\label{fig10}
\end{center}
\end{figure*}

\begin{figure*}
\begin{center}
\includegraphics[width=13cm,clip]{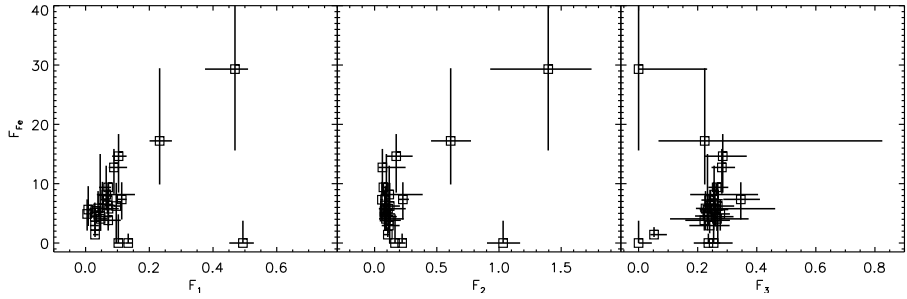}
\caption{Similar to  Fig. 10, but for the
long dips (10122-dip1, 30081, 60024). The left (middle)
panel shows a correlation, and no significant
relationship has been noticed in the right one.}
\label{fig11}
\end{center}
\end{figure*}

\begin{figure*}
\begin{center}
\includegraphics[width=13cm,clip]{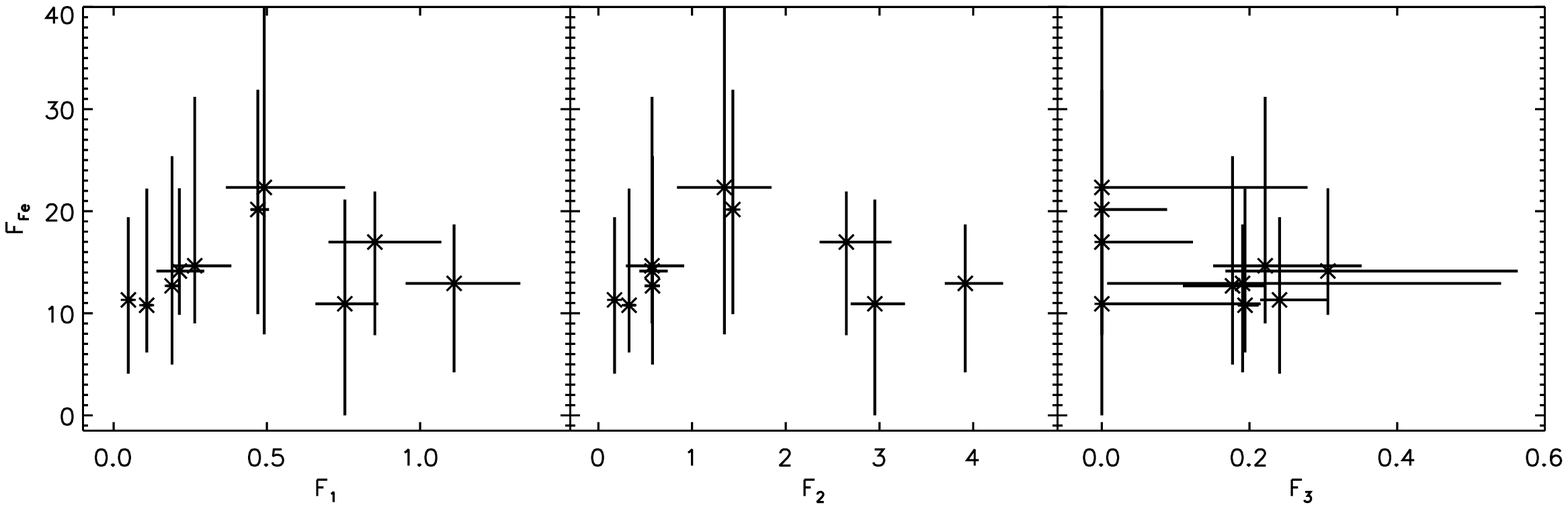}
\caption{Similar to  Fig. 10, but for the
short dips (10122-dip2, 30080).  The Fe line flux is almost
unchanged, within the errors, regarding to all continuum spectrum
fluxes.}
\label{fig12}
\end{center}
\end{figure*}

\section{Discussion}
\label{sect:discussion}

The binary system of Cir X-1 has a highly eccentric orbit. Possibly,
the NS may be occulted by the outer layers of its companion
atmosphere when the star moves close to the periastron.
The high eccentricity of the orbit might produce large tidal
interactions at this phase, in turn, which perhaps expand and/or
deform the  accretion stream, or maybe form a bulge in the outer accretion disk.
The excess of these matters may influence on the observed emission
of X-ray source, producing a partial covering of the energy
spectrum of the source (e.g., Inoue 1989; Iaria et al. 2001).

The Fe K$\alpha$ line emission at 6.4$\sim$6.7\,keV is a
common feature in LMXBs, and its origin has been an interesting
topic. Generally, there are four possible sites  for the Fe
K$\alpha$ emissions: the accretion disk, ADC, the source itself, or
the scattering medium (e.g., Smale et al. 1993; Shirey et al. 1999; Asai et al.2000).
The observed strong broad line at 6.4$\sim$6.7\,keV is
generally interpreted as arising from the ADC and illuminated outer
disk (e.g., White et al. 1985; Hirano et al. 1987; Asai et al.
2000), but it is difficult to obtain both the observed line
intensity and width from such a corona.

For Cir X-1, the results of Shirey et al. (1999) show that
the Fe K$\alpha$ line flux is constant and associated with the faint 
component during short dips,
implying that the Fe K$\alpha$ comes from the scattering medium far
away from the central region.
After investigating  the long dip of Cir X-1, Ding et al. (2006a,b)
propose that Fe line emission is from the disk.
Our results show that the flux of iron line emission changes
obviously during the long dips (see OBSID 30081 of Table 3) and is positively
correlated with the fluxes of  blackbody(1) and blackbody(2).
This token implies that the line could not be associated with the
soft component unaffected by the partial covering,
and instead, it is affected by the partial covering and could come 
from the region close to the central object.


For short dips, our result shows that the flux of the Fe
emission line is almost constant.
There are two possible interpretations for this constant Fe line emission
flux: (1) for the short dip the obscuring
matter could be of a small size, and (2) the obscuring matter lies inner 
than the Fe line emission materials on the accretion disk. 
On the other hand, according to the fact that the normalization of
the Fe line is positively correlated with $N_{\rm H}$ (see Fig. 5),
we also considered that the Fe line could come from the accreting
matter, which suggests that the Fe line strength is related to
mass accretion rate. 
Dong et al. (2004) showed that the equivalent width of the Fe line was 
associated with column density, implying that the line could come from 
the cold matter invoked in an absorbed model for active galactic nuclei (AGNs) NGC 2110 and NGC 7582.
However, Asai et al. (2000) analyzed the Fe
K$\alpha$ emission lines in the spectra of 20 LMXBs using
{\sl ASCA} data, and found that the centroid energies of the Fe
lines are almost independent of the source luminosity, and the
detected equivalent width of the line also does not show a clear
correlation with the source categories. 
Although the physical parameters of determining the distinction of
these categories has not been figured out,  the inclination angle,
mass accretion rate, and the stellar magnetic field are the most
relevant quantities. Thus, Asai et al. (2000) proposed that these 
parameters may not play a key role in determining the Fe line strength.
Unlike the conclusions of Asai et al. (2000), our results favor
that the Fe emission line might be correlated with the accreting
matter and come from the inner accretion disk.

Nevertheless, we also noticed some discussions about the origin of
the Fe emission line of other objects. For some black hole
candidates, there is evidence that the observed Fe K$\alpha$ line
originates in the innermost part of the accretion disk, close to the
black hole (e.g., Fabian et al. 1989; Cui et al. 1998; Zycki et al. 1999).
In addition, for AGNs, the higher
resolution spectra available with the {\sl ASCA} appear to indicate that
the line profiles are broad and asymmetrically skewed to lower
energies, which is interpreted as evidence that the majority of the
line emission origin from the inner accretion disk around the
massive black hole (Nandra et al. 1997; Reynolds
\& Begelman 1997). 
Nandra et al. (2000) presented a spectral analysis of the Seyfert 1
galaxy NGC 7469 with {\sl RXTE}, and found a significant correlation
between the 2-10\,keV flux and the 6.4\,keV Fe K$\alpha$
line, suggesting that the line emission comes from very nearby
central regions.  
 Our results imply that the Fe line emission region
on the accretion disk of an X-ray binary may be located at the 
same relative region as on the disk of an AGN.

\section{Conclusion}
\label{sect:conclusion}

In this paper, we analyzed the X-ray dip spectra of Cir X-1. The
spectra can be well fitted with two blackbody components with the
partial covering plus a third blackbody which is unaffected by
partial covering.
According to the spectral fitting results, the equivalent blackbody
emission radii of the first (T$\sim$2\,keV), second (T$\sim$1\,keV) and third
(T = 0.5$\sim$0.6\,keV) component are, respectively, 4-7\,km, 16-30\,km, 30-50\,km.
The fluxes of the first and  second components are closely
correlated, indicating that they represent emission from the inner regions.
The emitting region of the
coolest (3rd) component might come from the outer accretion disk.

A 6.4-6.7\,keV Fe emission line was detected in the  dip
spectra.  
During the long dips, the Fe line flux and fluxes
of the first and  second components are correlated, and
there is no significant correlation between the Fe line
flux and the flux of the  third component, which suggests the Fe
emission line comes from the region very close to the central
object and is affected by the covering matter. During the short dip, 
the Fe line flux is almost
constant within the errors for all continuum spectra fluxes, which
might be due to the fact that the line emission region lie in 
the outer of the obscuring matter or the size of the obscuring matter
is much smaller than that of the Fe line emitting region.

\begin{acknowledgements}
We would like to thank T.P. Li and S. N. Zhang for useful discussions.
This research has made use of data obtained through the
High Energy Astrophysics Science Archive Research Center
Online Service, provided by the NASA/Goddard Space
Flight Center. We acknowledge the {\sl RXTE} data teams at
NASA/GSFC for their help. This work is subsidized by the
Special Funds for Major State Basic Research Projects and
by the National Natural Science Foundation of China.
We are very grateful for the critic comments from the anonymous referee
 that greatly improve the  quality of the paper.
\end{acknowledgements}

\label{lastpage}
\end{document}